\documentstyle[preprint,prabib,aps]{revtex}
\begin{document}
\draft
\def\gappeq{\mathrel{\rlap {\raise.5ex\hbox{$>$}}
{\lower.5ex\hbox{$\sim$}}}}
 
\def\lappeq{\mathrel{\rlap{\raise.5ex\hbox{$<$}}
{\lower.5ex\hbox{$\sim$}}}}
 
\def \gsim{\lower.8ex\hbox{$\sim$}\kern-.75em\raise.45ex\hbox{$>$}\;}
\def \lsim{\lower.8ex\hbox{$\sim$}\kern-.8em\raise.45ex\hbox{$<$}\;}
 
                      \def\ltsima{$\; \buildrel < \over \sim \;$}
                      \def\simlt{\lower.5ex\hbox{\ltsima}}
                      \def\rtsima{$\; \buildrel > \over \sim \;$}
                      \def\simrt{\lower.5ex\hbox{\rtsima}}

\title{
Hadronic sizes and observables in high-energy scattering } 
 
\author{Erasmo Ferreira}
\address{
Instituto de F\'\i sica, Universidade Federal do Rio de Janeiro\\
 Rio de Janeiro 21945-970, RJ, Brazil\\}
 
\author{Fl\'avio Pereira}
\address{ Observat\'orio Nacional, CNPq       \\
 Rio de Janeiro 20921-400, RJ, Brazil\\}

 
\maketitle
 
\begin{abstract}
     The functional dependence of the high-energy observables
of total cross section and slope parameter on the sizes of 
the colliding hadrons predicted by the model of the stochastic 
vacuum and the corresponding relations used in the geometric model 
of Povh and H\"ufner are confronted with the experimental data.
The existence of a universal term in the expression for the 
slope, due purely to vacuum effects, independent of the energy
and of the particular hadronic system, is investigated.
\end{abstract}
\bigskip
 PACS Numbers~:~12.38 Lg , 13.85 Dz , 13.85 Lg~.

 \newpage

{\bf 1. The Model of the Stochastic Vacuum and the Geometric Models }
\bigskip

Diffractive high energy scattering is largely determined by the
nonperturbative regime of QCD. The extended character of the 
interaction, involving correlation properties of the gauge field, 
determines the phenomenological properties of the observables, 
which are fixed by the sizes and global structures of the colliding 
systems, rather than by the number of their pointlike constituents 
and their couplings. 
These features have led to models of geometric nature
for high-energy scattering \cite{RRB5A,RRB5}, which give  
natural account of the relations between total cross sections of 
different hadronic systems  and, through 
hadronic form factors appropriately introduced, describe 
the shapes of the diffractive peaks. QCD must provide the fundamental 
framework  in which 
this phenomenology should arise naturally, and efforts have been made in 
this direction. Some treatments based on perturbative QCD  have   
also lead to dependence of the observables on hadronic sizes
\cite{RRB5B}.
   
  A nonperturbative QCD description of the main features of 
high-energy scattering is given by the model of the stochastic 
vacuum  \cite{RRG3,RRG3A}, which  combines QCD quantities (gluon 
condensate and corrrelation length) and hadronic sizes in 
an eikonal framework, 
leading to a unified description of  the data for different 
hadronic systems. The calculations lead to  definite size dependence of  
the observables of total cross section  $\sigma^T$ and forward slope 
parameter B, and provide a dynamical scheme that explains
the intuitive and Regge-based geometric treatments. 
In the present work we explore further the results of the model 
for the  pp, $\bar {\rm p}{\rm p}$ and other hadronic 
systems, and compare them with those from the 
geometric model of B.Povh and J.H\"ufner \cite{RRB5}, confronting 
all predictions with the experimental data.

In the calculation with the model of the stochastic vacuum
the hadronic structures enter in the form of transverse wavefunctions
(two-dimensional wavefunctions in the plane perpendicular to the
direction of the colliding hadrons). Taking into account the results 
of the previous analysis of different hadronic systems  \cite{RRG3}  we
here only consider for the proton a diquark structure  where, with respect 
to the relevant color degrees of freedom,  the proton
is described as a meson, in which the diquark replaces the antiquark.
Thus we can treat on equal footing meson-meson, meson-baryon 
and baryon-baryon scattering. Other structures for the baryons 
have been explored \cite{RRG3,RRG3B} in investigations with the model 
of the stochastic vacuum, and the similarity of results obtained 
with different structures demonstrates the role of the extended 
nature of the nonperturbative dynamics.
  
           For the hadron transverse wavefunction we take the 
simple ansatz
\begin{equation}
     \psi _{H} (R)=  \sqrt {2/\pi}\frac {1}{S_H} \exp{(-R^{2}/S_H^{2})}~,
\label{4A12}
\end{equation}
     where $S_H$  is a parameter for the hadron size.
The dimensionless scattering amplitude $T_{H_1H_2}$  is given in terms 
of the dimensionless profile function $\widehat J_{H_1H_2}$ for 
hadron-hadron scattering by 
\begin{equation}
 T_{H_1 H_2} = i s [\langle g^2 FF\rangle a^4]^2 a^{2}\int d^2 \vec b~
          \exp{(i \vec q\cdot \vec b )}~\widehat J_{H_1 H_2}
 (\vec b,S_1,S_2) ~,
 \label{4A14}
 \end{equation}
  where the impact parameter vector  $\vec b$  and the hadron sizes
  $S_1$ , $S_2$ are here written in units of the
     correlation length $a$, and $ \vec q $ is the momentum
     transfer projected on the transverse plane, in units of $1/a$,
     so that the momentum transfer squared is $t=-|\vec q|^2/a^2$. 
   For short, from now on we write $J(b)$ or $J(b/a)$  to represent
$ \widehat J_{H_1 H_2} (\vec b,S_1,S_2)$.
        The normalization of $T_{H_1 H_2}$ is such that total
and differential cross sections are given by 
\begin{eqnarray}
 \sigma^T  = \frac{1}{s}~\hbox{\rm Im}~T_{H_1 H_2} ~~~,~~~
    \frac {d\sigma^{e\ell}}{dt} =
             \frac {1}{16\pi s^2}~\vert T_{H_1 H_2}\vert^2  ~.
\label{4A15}
\end{eqnarray}

        The observables are written in terms of 
 dimensionless moments of the profile function 
(as before, with $b$ in units of the correlation length $a$)
\begin{equation}
              I_k = \int d^2\vec b~b^k~ J(b) ~~,~k=0, 1, 2, ...
\label{4A18}
\end{equation}
which depend only on $ S_1/a$, $ S_2/a$, and the Fourier-Bessel transform
\begin{equation}
              I(t) = \int d^2\vec b~ J_0 (b a \sqrt{|t|})~ J(b) ~,
\label{4A19}
\end{equation}
where $J_0(b a \sqrt{|t|})$ is the zeroth--order Bessel function.
Then 
\begin{equation}
T_{H_1 H_2} = i s[\langle g^2 FF\rangle a^4]^2 a^{2} I(t)~. 
\label{4A19A}
\end{equation}
  Since  $J(b)$ is real, $\sigma^T$ and the
slope parameter B are written 
\begin{eqnarray}
            \sigma^T= I_0~[\langle g^2 FF \rangle a^4]^2 a^{2} ~~,~~
   B = \frac {d}{dt} \biggl( \ln \frac {d\sigma^{e\ell}}{dt} \biggr)
   \bigg\vert_{t=0} ~=\frac{1}{2}~ \frac{I_2}{I_0} ~a^2 ~ \equiv K a^2~.
\label {4A21}
\end{eqnarray}

   It is important to observe that these results conveniently factorize
the dimensionless QCD strength $ \langle g^2 FF \rangle a^4$ in the 
expressions for the observables. The correlation length $a$, which is 
an intrinsic parameter of the correlation function of the QCD field, 
appears as the natural length scale for the 
observables and for the geometric aspects of the interaction. 
 These aspects are concentrated on the quantities $I_0(S_1/a,S_2/a)$ and
$I_2(S_1/a,S_2/a)$, which depend on the hadronic structures. 
 These quantities are mainly determined by the
values of the profile functions  in the range of impact parameters 
up to about 2.5 fm.
$\sigma^T$ measures the strength, while the slope B has the strength 
cancelled out and is only related 
to the hadron geometry. 
The explicit formula for the slope is
\begin{equation}
B= \frac{1}{2}\frac{ \int d^2\vec b~b^2~ J(b)}
          {\int d^2\vec b~ J(b)}~a^2= \frac{1}{2}~\langle b^2 \rangle a^2 ~,
\label{4A21A}
\end{equation}
where it is seen as related to the average value of the square of the 
impact parameter in the collision, with J(b) as weigth function.
We recall that here b is dimensionless and that  $\langle b^2 \rangle$ 
depends on the hadronic sizes.

\bigskip

{\bf 2. pp and $\bar {\rm p}$p systems }

 We first discuss  pp and $\bar{\rm p}{\rm p}$ systems, with $S_1=S_2=S$. 
    The curves for  $I_0= \sigma^T /~[\langle g^2 FF \rangle^2 a^{10}] $~ 
and $ K= B/a^2 $ can be parametrized as simple powers of $S/a$ 
with good accuracy, the convenient expressions being   
\begin{eqnarray}
I_0= \alpha \bigg(\frac{S}{a}\bigg)^{\beta}~~~,~~~ 
          K=\eta +\gamma \bigg(\frac{S}{a}\bigg)^{\delta}~.
\label{4A27}
\end{eqnarray}
The values of the parameters result from integrations over correlation 
functions \cite{RRG3}, 
are intrinsic to the model of the stochastic vacuum, and do not 
contain any dependence on  experimental quantities. For the present 
purpose of analysis of data in a limited energy range, and for easier 
comparison to the geometric model we take their values as $\eta=2.03$, 
$\beta=8/3$, $\gamma=3/8$, $\delta=2$ , $\delta/\beta=3/4~$, 
$\alpha=0.76\times 10^{-2}$ . 

     The proton radius can be eliminated from
 Eqs. (\ref{4A21}) and  (\ref{4A27}), and we obtain 
a relation between the observables $\sigma^T$ and $B$ at a given energy  
\begin{equation}
(B - \eta a^2)= \frac{a^2}{[<g^2 FF>a^4]^{2\delta/\beta}}
           \frac{\gamma}{\alpha^{\delta/\beta}}
\bigg(\frac{\sigma^T_{\rm pom}}{a^2}\bigg)^{\delta/\beta}~.
\label{5A29}
\end{equation}
  The two QCD parameters, $\langle g^2 FF\rangle$ and  $a$, 
can be determined using this expression and the
  experimental data for $\sigma^T$ and B at two different energies.

  The  available data on $\sigma^T$ and B  in 
pp and $\bar {\rm p}$p scattering at high energies  
consist mainly \cite{RRDAT1} of ISR (CERN) 
measurements at energies ranging from $\sqrt{s}=23$ GeV to 
$\sqrt{s}=63$ GeV, of the $\sqrt{s}=541 - 546$  GeV  measurements
in CERN SPS and in Fermilab, and of the $\sqrt{s}=1800$ GeV
data from the E-710 Fermilab experiment. 
The Fermilab CDF  measurements \cite{RRDAT3} at $\sqrt{s}=1800$ GeV
seem discrepant with the E-710 experiment at the same energy and are not 
used here. 

  Since we are here concerned with nonperturbative contributions
only, at the ISR energies we take for total cross sections the 
values given by the Donnachie and Landshoff parametrization \cite{RRB2}
for the pomeron-exchange contribution
\begin{equation}
\sigma^T_{\rm pom}({\rm pp,\bar pp} )=( 21.70~{\rm mb})~s^{0.0808}~,
\label{5A1}
\end{equation}
and for values of the slope we take those of the pp system (not 
those of the  $\bar{\rm p}{\rm p}$ data). Using as input the 
data for the highest energies (541 and 1800 GeV), where 
the process is essentially 
nonperturbative and no separation is needed, we obtain \cite{RRG3A} 
\begin{equation}
a=0.32\pm 0.01~ {\rm fm}~,
    ~<g^2 FF>a^4 = 18.7\pm 0.4~,~ <g^2 FF>=2.7\pm 0.1~{\rm GeV}^4~.
\label{5A35A}
\end{equation}

The relation between the experimental values of the two observables 
is well represented at all energies from 23.5 to 1800 GeV with the form
\begin{equation}
B=B_\Delta + C_\Delta  (\sigma^T)^\Delta~.
\label{1A1}
\end{equation} 
We use this expression with B and B$_\Delta$ in GeV$^2$, $\sigma^T$ in 
mb and $C_\Delta$ in mixed units.
This form is similar to Eq. (\ref{5A29}), 
with an obvious correspondence of parameters.  In our calculation  with 
the model of the stochastic vacuum  the exponent 
$\Delta=\delta/\beta $ does not depend on QCD quantities and is 
equal to about 0.75. 
 
 Following ideas that relate hadron-hadron 
scattering to the shape and size of the colliding hadrons, 
B.Povh and J.H\"ufner \cite{RRB5} show that 
the combination of Regge amplitudes with electromagnetic form factors 
relates the slope parameter and the sum of 
the squares of the radii of the colliding hadrons, and  
write for Hp (hadron-proton) scattering
     \begin{equation}
     B_{\rm{Hp}}=R_{\rm p}^2+R_{\rm H}^2~,
\label{6A1}
\end{equation}
which is to be considered as a definition of effective hadronic radii. 
 Observing the behavior of the experimental points in a plot of  the 
observables $\sigma^T$ and B against each other, they suggest that the
dependence  of the total cross section on the radii is
\begin{equation}
\sigma^T_{\rm {Hp}}= g R_{\rm {H}}^2R_{\rm{ p} }^2~.
\label{6A2}
\end{equation}
Introducing electromagnetic form factors to reproduce the shape of the elastic
differential cross section, B.Povh and J.H\"ufner  write the relation of these 
hadronic radii to the electromagnetic radii as $<r^2_{\rm em}>=3R^2$.  
Deviations from these simple formulae occurring at low energies show 
that they should be used only for $\sqrt{s}\geq 20~{\rm GeV}$.

    For proton-proton scattering, with $R_{\rm H}=R_{\rm p}$, 
Eqs. (\ref{6A1}) and (\ref{6A2}) lead to 
\begin{equation}
B= C_{1/2}~(\sigma^T)^{1/2}~,
\label{6A3}
\end{equation}
which is of the form of Eq. (\ref{1A1}) with $\Delta=1/2$ and $B_\Delta=0$, 
 and it is 
remarkable that, using as input the data at 541 and 1800 GeV, 
 one obtains with $\Delta=1/2$ the same value $B_\Delta=0$.

   In Fig. 1 we show the description of the data through Eq. (\ref{1A1}), 
using for $\Delta$ the  values 1, 0.75 and 0.5. 
   The case $\Delta=1$ is included in Fig. 1 for a numerical reference, 
although we do not refer to any  model suggesting it. 
The vertical axis represents the constant C$_\Delta$, which,
together with B$_\Delta$, is fixed in each case by the input data at the  
energies 541 and 1800 GeV. Then the five ISR data 
points are considered as parameter-free 
predictions, and 
we calculate a $\chi^2$ value representing the observed deviations.
 In the horizontal axis we mark the 
energy, which here works just as an external label used 
to spread the information in the plot. There are no free parameters, 
since B$_\Delta$ and C$_\Delta$ are fixed by the input data.    
The values of $\chi^2$ are also shown, and, although
$\Delta=0.75$ is favoured, we cannot say that the differences are 
statistically meaningful. However, the model of the stochastic vacuum  
gives precise meaning to the parameters  B$_\Delta$ and C$_\Delta$ in 
terms of QCD quantities, successfully predicts $\Delta=0.75$ ,
and introduces  hadronic sizes in definite form, as 
parameters accounting for the extensions of the wavefunctions.
  It is remarkable the presence in this case of a 
bounding minimum $B_\Delta$ (equal to $\eta a^2$) for the slope, which 
is the same for all hadron-hadron systems. 

\vspace{13cm}
{\small
{\bf Fig. 1~-~}Test of the parameters of Eq. (\ref{1A1}), comparing 
different models, using the experimental quantities of 
the pp and $\bar{\rm p}{\rm p}$ systems. For energies up to 62.3 GeV
the values of $\sigma^T$  are  given by the 
parametrization  $\sigma^T_{\rm pom}=(21.70~{\rm mb}) s^{0.0808}$  
and the values of B are those of the pp data. 
The values of  $B_\Delta$ and $C_\Delta$ are obtained with the 541 
GeV and 1800 GeV  data  as inputs.
The horizontal lines represent the constant $C_\Delta$ 
 with the choices $\Delta$=1, 0.75 (model of the stochastic vacuum) 
 and 0.5 (geometric model). $\chi^2$ represents the average deviation 
of   the five ISR points from the constant line. 
  }
\bigskip

The description given in Fig. 1 covers all pp and $\bar{\rm p}$p data. 
Extrapolating Eq. (\ref{1A1})
to higher energies (e.g. LHC energies), where the total
cross sections may be about 100 mb, we find  a small, but 
hopefully measurable, difference in the values of the slope, with B 
higher by 0.2 GeV$^{-2}$ for $\Delta=3/4$ , compared with the 
$\Delta=1/2$ case. 

The proton radius presents a slow increase with the energy, taking 
values about the electromagnetic radius. In the case of the model of 
the stochastic vacuum,  
where the radius enters as a parameter of the wavefunction, the
energy dependence of the radius can be parametrized in the form
\cite{RRG3A}
\begin{eqnarray}
    S_{\rm p}(s)   &=& 0.671 + 0.057 \log \sqrt{s}~~({\rm fm})~~~~~~~~~~(a)~, 
                     \nonumber \\
{\rm or} ~~~~
   S_{\rm p}(s)&=& 0.572 + 0.123~ [\log \sqrt{s}]^{0.75}~~({\rm fm})~~~~(b)~.
\label{5A19B}
\end{eqnarray}
 In the geometric model Eq. (\ref{6A2})
requires a power 1/2 in the logarithm 
in order to yield  a $\log^2 s$ dependence in the cross section.

  Parametrizations (a) and (b) predict for $\sqrt{s}= 14$ TeV values of 
the proton radius 1.215 fm and 1.240 fm respectively, which are about
40\% higher than the electromagnetic radius. The pp cross section at 
this energy is predicted as (a) 95.5 mb and (b) 100.8 mb. These values 
are in good agreement  with the results of the Akeno collaboration
\cite{cosmic}.

The value of  g is obtained by B.Povh and J.H\"ufner, on the basis of 
Hp data, as g=75 fm$^{-2}$. In Fig. 1 we show that 
$C_{1/2}\approx 2 ~ {\rm GeV}^{-2} ~ {\rm mb}^{-1/2}$, 
corresponding to $ g=(2/C_{1/2})^2 \times 0.1/(0.197)^4=66.4~{\rm fm}^{-2}$.
This means a fair simultaneous description of pp and pH systems,
which is also achieved in the calculations with the model of the 
stochastic vacuum, where the interaction strength and the hadronic 
radii appear as fundamental quantities. An important phenomenological 
difference between the two approaches rests in the existence of a finite 
universal minimum value $B_\Delta$ for the slope, which may
be tested in systems where  smaller  hadrons  collide with the proton, 
as we discuss below. 

  Eqs. (\ref{6A1}), (\ref{6A2}) and (\ref{6A3}) for pp and 
$\bar {\rm p}{\rm p}$ scattering 
 can be built from a profile function of simple Gaussian shape
\begin{equation}
J(b)= \frac{gR_{\rm p}^2}{4\pi}~e^{-b^2/4R_{\rm p}^2}~,
\label{6A4}
\end{equation}
where $R_p(s)$ has an energy dependence, and where the normalization
for J is chosen appropriately. However, two Gaussians are needed to 
describe the data for $t\neq 0$.

{\bf 3. Hadron-Proton Systems }
  
 We now consider  other systems of hadrons colliding at high
energies. In the treatment of the pp system we are constrained  by 
$\sqrt{s}\geq 20~{ \rm GeV}$, and cannot observe clearly  the effect of 
the minimum slope $B_\Delta$ . The contribution of this term could be
better observed in Hp  systems, where H represents hadrons of small size. 
We must remark that, since we deal with radii which are 
energy dependent quantities,  we must  compare 
different hadronic systems at the same center-of-mass energy.

   The parametrization of the results obtained with the model of the 
stochastic vacuum for general  Hp systems is 
\begin{eqnarray}
\sigma^T_{\rm pom} = I_0~[\langle g^2 FF \rangle a^4]^2 a^{2}
   =[\langle g^2 FF \rangle a^4]^2 a^{2} 
     \alpha \bigg(\frac{S_{\rm p}}{a} \frac{S_{\rm H}}{a}\bigg)^{\beta/2}~,
\label{4A21F}
\end{eqnarray}
and 
 \begin{eqnarray}
   B = \frac{1}{2}~ \frac{I_2}{I_0} ~a^2 ~ 
   = \eta~a^2 + \frac{1}{2} \gamma \bigg(S_{\rm p}^2+S_{\rm H}^2\bigg)~.
\label {4A21G}
\end{eqnarray}
With $a=0.32~{\rm fm}$, we have $\eta a^2=5.38~{\rm GeV}^{-2}~$.
  In the treatment of B.Povh and J.H\"ufner the corresponding 
relations are given by Eqs. (\ref{6A1}) and (\ref{6A2}). 
   
    In order to compare the models, it is important to eliminate 
the influence of specific values of radii, since they have 
different definitions. Thanks to the convenient factorization in 
the final expressions, we may actually build relations involving only
the observables, or involving only the ratios of radii, which we 
may assume to follow the ratios of electromagnetic radii. We thus have
for $\sigma_{\rm Hp}/\sigma_{\rm H'p}$ the ratios 
$(r_{\rm H}/r_{\rm H'})^{4/3}$
and $(r_{\rm H}/r_{\rm H'})^2$ in the stochastic vacuum and  geometric models 
 respectively. 
Entering with the known  values \cite{RADII}  for the radii of the 
proton ($0.862\pm 0.012~{\rm fm} $), of the pion ($0.66\pm 0.01~{\rm fm}$) 
and of the kaon ($0.58\pm 0.04~{\rm fm}$) we obtain the results 
shown in Table I. The experimental ratio refers to the pomeron
exhange contribution, taken from the parametrization of Donnachie
and Landshoff. We observe that the value  2/3 given for the 
ratio $\sigma_{\pi p}/\sigma_{pp}$ by the quark additivity rule is 
here obtained as a simple consequence of the sizes of the hadrons.
Also the ratio $\sigma_{Kp}/\sigma_{\pi p}$ is consistently obtained 
with a value close to the data, without need for  different couplings
of the pomeron to strange and non-strange quarks, as must be the case 
with quark additivity rules. 
Since the analysis of the proton structure in HERA (DESY) shows that 
the proton is better characterized as a {\it sea} rather than as a 
{\it valence}  
structure, the explanation of the high-energy phenomenology through 
the hadronic sizes is more legitimate. The factorization 
relation $\sigma_{\pi\pi}=\sigma_{\pi p}^2 /\sigma_{\rm pp}$ is 
identically satisfied in both cases considered here. 

\bigskip  
{\bf Table I -} {\small  Ratios of the pomeron exchange contributions to  
total cross sections for different hadronic systems. The experimental values 
are taken from the parametrization of  Donnachie and Landshoff.}

   \begin{center}
  \begin{tabular}{|c|c|c|c|}  \hline
  Cross section                  & stochastic     & geometric  & Experimental \\
  ratios                          & vacuum        & model      & values \\ \hline
   $\sigma_{p\pi}/\sigma_{pp}$    & $0.69\pm 0.02$ & $0.59 \pm 0.02$ & 0.63\\ 
   $\sigma_{p K }/\sigma_{p\pi }$ & $0.83\pm 0.08$ & $0.77 \pm 0.08$ & 0.87\\
  \hline
  \end{tabular}
  \end{center}
     \vspace{1.0cm}

    Considering all Hp systems at a given energy, 
Eqs. (\ref{4A21F}) and (\ref{4A21G}) lead  
to a nonzero minimum possible value for the slope, given by 
\begin{eqnarray}
B_{\rm Hp}^{\rm min}(s)= \eta a^2 + \frac{\gamma}{2} S_{\rm p}^2
   =\frac{1}{2}\eta a^2 + \frac{1}{2}B_{\rm pp}(s)
= 2.69~ {\rm GeV}^{-2} + \frac{1}{2}B_{\rm pp}(s)~.
\label{4A21I}
\end{eqnarray}
The existence of  this minimum 
slope that can be observed in the scattering of any hadron 
by a proton is characteristic of the model of
the stochastic vacuum. To relate the observables for different 
Hp systems at a given energy, we call 
$G=\alpha [\langle g^2 FF \rangle a^4]^2 a^{2}$ and write 
\begin{eqnarray}
&&\frac{B_{\rm pp}-B_{\rm Hp}}
{\sigma_{\rm pp}^{4/\beta}-\sigma_{\rm Hp}^{4/\beta}}=
\frac{(\gamma/2)(S_p^2-S_H^2)} 
{G^{4/\beta}(S_p^2/a^2)^2-G^{4/\beta}(S_pS_H/a^2)^2} 
=\frac{(\gamma/2)~a^2}{G^{2/\beta}\sigma^{2/\beta}_{\rm pp}}~.
\label{4A21J}
\end{eqnarray}
The last quantity is fixed, for a given energy. As we go from
a hadron H to another, we obtain in a 
plot of $B_{\rm Hp}$ against $\sigma_{\rm Hp}$
a line from the point 
representing the observables of the pp system to
the limit point  $\sigma=0$, $B=B^{\rm min}$ 
given by Eq. (\ref{4A21I}). With $\beta=8/3$ we have 
\begin{equation}
B_{\rm Hp}=B_{\rm Hp}^{\rm min}+\frac {B_{\rm pp}-B_{\rm Hp}^{\rm min}}
            {\sigma_{\rm pp}^{1.5}}~\sigma_{\rm Hp}^{1.5}~.
\label{4A21K}
\end{equation}
Using the data for the pp system at 23.5 GeV 
we obtain $B_{\rm Hp}=8.59 + 0.014775 ~ \sigma_{H{\rm p}}^{1.5}$ .
This plot is shown in Fig. 2, together with data of the pp, $\pi$p
and Kp systems \cite{RRDAT4}
 at $\sqrt{s}\approx 20 ~ {\rm GeV}$. The limit point 
is shown inside a square window in the figure.

 In the case of the geometric model we have 
\begin{eqnarray}
B_{\rm Hp}^{\rm min}(s)= R_{\rm p}^2= \frac{1}{2}B_{\rm pp}(s)~
\label{4B21I}
\end{eqnarray}
  and a straight line in a plot of $B_{\rm Hp}$ against $\sigma_{\rm Hp}$

\begin{eqnarray}
&&\frac{B_{\rm pp}-B_{\rm Hp}}
{\sigma_{\rm pp}-\sigma_{\rm Hp}}=\frac{1}{(\sigma_{\rm pp} g)^{1/2}}~.
\label{4B21J}
\end{eqnarray}
$g$ is fixed by pure pp data putting $\sigma_{\rm Hp}=0$, 
$B_{\rm Hp}=B_{\rm pp}/2$, and using Eq. (\ref{6A3}). Then 
\begin{equation}
B_{\rm Hp}=\frac{1}{2} B_{\rm pp}~
         [1+\frac{\sigma_{\rm Hp}}{\sigma_{\rm pp}}]~.
\label{4B21K}  
\end{equation}

Comparing the two lines shown in Fig. 2 we may tell, for now with 
some subjective 
judgement, if either model describes better the data.

\vspace*{13cm}
{ \small
{\bf Fig. 2~-~}Observables for different hadronic systems at 20 GeV.
The straight line is the prediction of the geometric model, 
the square window showing the minimum value for the slope predicted 
for small hadrons colliding  with protons at this energy. 
The upper curve and the upper square window represent the 
predictions of the model of the stochastic vacuum. 
}
\bigskip

The model of the stochastic vacuum predicts that the
slope B for the $\pi\pi$ system at about $\sqrt{s}\approx 20$ GeV is 
$ B_{\pi\pi}= \eta a^2+ \gamma~ 2~ S_{\pi}^2\approx 9.6~ {\rm GeV}^{-2}~,$
while the geometrical model predicts 
$ B_{\pi\pi}= (2/3)~ r_{\pi}^2\approx 7.5~ {\rm GeV}^{-2}~.$
This is not a trivial difference, as it tests the 
contribution of a nonperturbative QCD effect.

\bigskip

{\bf 4. Conclusions }

The results presented in this work exhibit the simplicity of the 
connections between hadronic high-energy  observables determined 
by the  hadronic sizes and stress the importance of geometric 
relations as indicative of properties of nonperturbative QCD dynamics. 
 
Both the model of the
stochastic vacuum  and the geometric description of Povh and H\"ufner
give fair account of the present data, although it seems to us that 
the relations provenient of the QCD calculation are more accurate.
The differences between the two descriptions 
are important and interesting, and 
once they can be fully tested by the data, may become crucial.
The existence or not of the universal term $\frac{1}{2}\eta a^2$,
of unique value for all energies, representing a pure 
nonperturbative QCD 
contribution to hadronic scattering, is a question of fundamental
importance. 
Direct hadronic data on hadronic systems with small mesons, such 
as $\phi$p and $\psi$p would be very interesting for the 
study of nonperturbative QCD effects. Hopefully  the $\phi$ factory 
in Frascati will create the opportunity for these studies.

While the geometric relations of Povh and H\"ufner are basicaly 
empirical, the
quantitative details (the form  of the functional relations
and the intrinsic values of parameters) in the predictions of the model 
of the stochastic model determine fundamental QCD quantities 
using only a small amount of data. The model explains the
energy dependence of the observables in terms of the energy variation
of the hadronic sizes, and relates experimental 
quantities for different hadronic systems, exhibiting properties of the
extended nature of the interaction, which is determined by the 
structure of the QCD vacuum. 



\newpage

\centerline {FIGURE AND TABLE CAPTIONS }

\bigskip

{\small
{\bf Fig. 1~-~}Test of the parameters of Eq. (\ref{1A1}), comparing 
different models, using the experimental quantities of 
the pp and $\bar{\rm p}{\rm p}$ systems. For energies up to 62.3 GeV
the values of $\sigma^T$  are  given by the 
parametrization  $\sigma^T_{\rm pom}=(21.70~{\rm mb}) s^{0.0808}$  
and the values of B are those of the pp data. 
The values of  $B_\Delta$ and $C_\Delta$ are obtained with the 541 
GeV and 1800 GeV  data  as inputs.
The horizontal lines represent the constant $C_\Delta$ 
 with the choices $\Delta$=1, 0.75 (model of the stochastic vacuum) 
 and 0.5 (geometric model). $\chi^2$ represents the average deviation 
of   the five ISR points from the constant line. 
  }
\bigskip

{ \small
{\bf Fig. 2~-~}Observables for different hadronic systems at 20 GeV.
The straight line is the prediction of the geometric model, 
the square window showing the minimum value for the slope predicted 
for small hadrons colliding  with protons at this energy. 
The upper curve and the upper square window represent the 
predictions of the model of the stochastic vacuum. 
}

\bigskip

\bigskip  
{\bf Table I -} {\small  Ratios of the pomeron exchange contributions to  
total cross sections for different hadronic systems. The experimental values 
are taken from the parametrization of  Donnachie and Landshoff.}

   \begin{center}
  \begin{tabular}{|c|c|c|c|}  \hline
  Cross section                  & stochastic     & geometric  & Experimental \\
  ratios                          & vacuum        & model      & values \\ \hline
   $\sigma_{p\pi}/\sigma_{pp}$    & $0.69\pm 0.02$ & $0.59 \pm 0.02$ & 0.63\\ 
   $\sigma_{p K }/\sigma_{p\pi }$ & $0.83\pm 0.08$ & $0.77 \pm 0.08$ & 0.87\\
  \hline
  \end{tabular}
  \end{center}
     \vspace{1.0cm}


\bigskip

\begin{thebibliography} {99}
\bibitem{RRB5A}C. Bourrely, J. Soffer and T.T. Wu, Nucl. Phys. 
{\bf B247}, 15 (1984);
 Phys. Rev. Lett. {\bf 54}, 757 (1985); Phys. Lett. {\bf B196}, 237
 (1987); J. Dias de Deus and P. Kroll,  Nuovo Cimento {\bf A37}, 67 (1977);  
 Acta Phys. Pol. {\bf B9}, 157 (1978); J. Phys. {\bf G9}, L81 (1983); 
P. Kroll, Z. Phys. {\bf C15}, 67 (1982); 
T.T. Chou and C.N. Yang, Phys. Rev. {\bf 170}, 1591 (1968); {\it ibid.}
                 {\bf D19}, 3268 (1979);
 Phys. Lett. {\bf B128}, 457 (1983); {\it ibid.} 
                  {\bf B244}, 113 (1990). 
 \bibitem{RRB5}B. Povh and J. H\"ufner, Phys. Rev. Lett.
          {\bf 58}, 1612 (1987); Phys. Lett. {\bf B215}, 772 (1988); 
   {\it ibid.} {\bf B245}, 653  (1990); 
           Phys. Rev. {\bf D46}, 990 (1992);  
          Zeit. Phys. {\bf C63}, 631 (1994).
\bibitem{RRB5B}J.F. Gunion and D.E. Soper, Phys. Rev. {\bf D15}, 2617 
     (1977); 
E. Levin and M.G. Ryskin, Sov. J. Nucl. Phys. {\bf 34}, 619 (1981).
\bibitem{RRG3}H.G. Dosch, E. Ferreira and A. Kr\"amer,  Phys. Lett.
          {\bf B289}, 153 (1992); {\it ibid.} {\bf B318}, 197 (1993);
     Phys. Rev. {\bf D50}, 1992 (1994).
\bibitem{RRG3A}E. Ferreira and F. Pereira, Phys. Rev. {\bf D55},130 (1997). 
\bibitem{RRG3B}H.G. Dosch and M. Rueter, Phys. Lett. {\bf B205}, 117 (1996). 
\bibitem{RRDAT1}Data on pp and $\bar {\rm p}{\rm p}$ systems. 
 (a) N. Amos {\it et al.}, Nucl. Phys. {\bf B262}, 689 (1985); 
 (b) R. Castaldi and G. Sanguinetti,  Ann. Rev. Nucl. Part. Sci.
         {\bf 35}, 351 (1985); 
 (c) C. Augier {\it et al.}, Phys. Lett. {\bf B316}, 448 (1993); 
 (d) M. Bozzo {\it et al.},  Phys. Lett. {\bf B147}, 392 (1984);   
    M. Bozzo {\it et al.}, {\it ibid.} {\bf B147}, 385 (1984); 
  (e) N. Amos {\it et al.}, Phys. Lett. {\bf B247}, 127 (1990); 
     Phys. Rev. Lett. {\bf 68}, 2433 (1992); 
\bibitem{RRDAT3}F. Abe {\it et al.},  Phys. Rev. {\bf D50}, 5550 (1994);
 {\it ibid.} {\bf D50}, 5518 (1994).
\bibitem{RRB2}A. Donnachie and P.V. Landshoff, Phys. Lett.
          {\bf B296}, 227 (1992).
\bibitem{RRDAT4}Data on slopes of $\pi$p and Kp systems.  
 (a) J.B. Burq {\it et al.}, Nucl. Phys. {\bf B217}, 285 (1983); 
 (b) A. Schiz {\it et al.},  Phys. Rev. {\bf D24}, 26 (1981);  
 (c) L.A. Fajardo {\it et al.}, Phys. Rev. {\bf D24}, 46 (1981); 
 (d) N. Adamus {\it et al.}, Phys. Lett. {\bf B186}, 223 (1987). 
\bibitem{RADII}Electromagnetic radii. 
(a) Proton :~ G.G. Simon {\it et al.},  Z. Naturforschung {\bf 35A}, 1 (1980); 
(b) Pion :~ S.R. Amendolia {\it et al.}, Nucl. Phys. {\bf B277}, 168 (1986); 
(c) Kaon :~ S.R. Amendolia {\it et al.}, Phys. Lett. {\bf B178}, 435 (1986).
\bibitem{cosmic}M. Honda {\it et al.},  Phys. Rev. Lett. {\bf 70}, 525 (1993).

 \end{thebibliography}
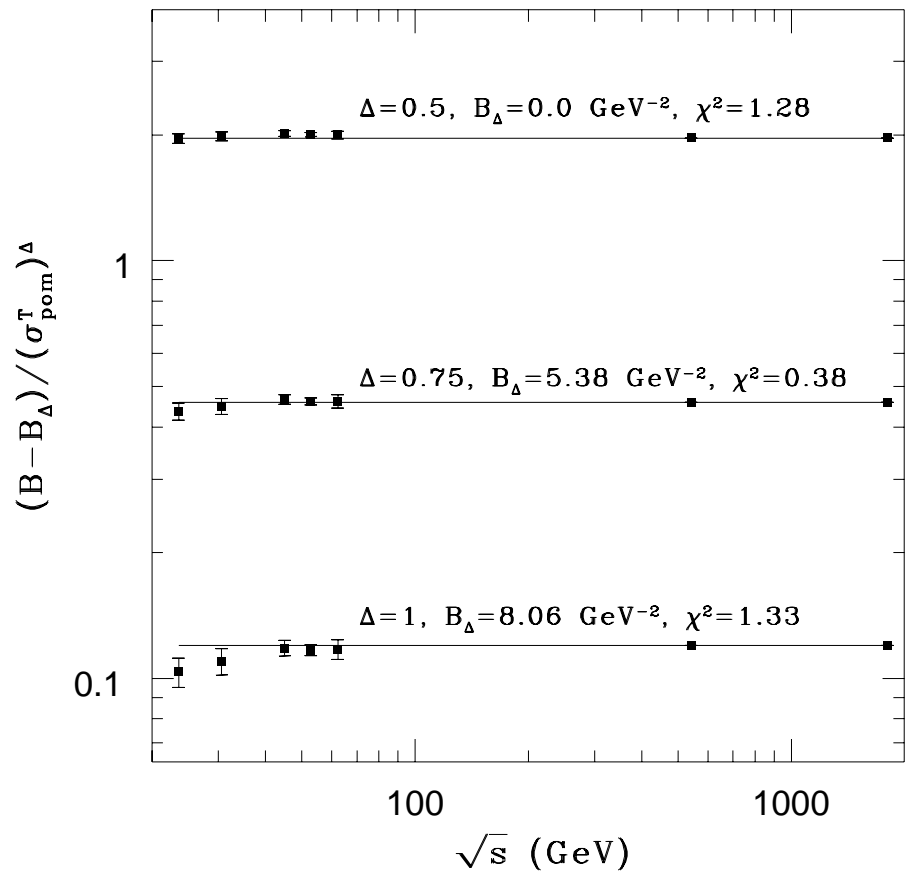
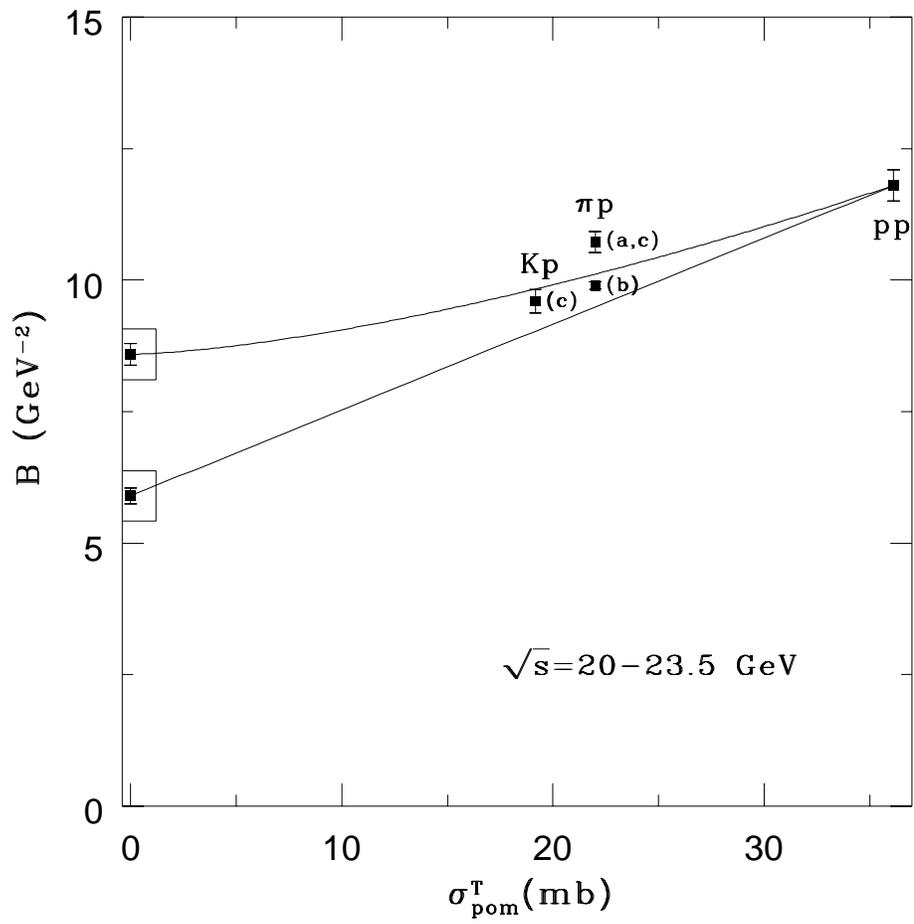
\end{document}